\documentclass{ws-p8-50x6-00}

\begin{document}

\title{Partition Function Zeros of\\ Aperiodic Ising Models}

\author{UWE GRIMM}

\address{Applied Mathematics Department, Faculty of Mathematics
and Computing,\\ The Open University, Walton Hall, 
Milton Keynes MK7 6AA, UK\\ 
E-mail: u.g.grimm@open.ac.uk}

\author{PRZEMYS{\L}AW REPETOWICZ}

\address{Department of Mathematics, Heriot-Watt University,
Riccarton,\\ Edinburgh EH14 4AS, UK\\
E-mail: p.repetowicz@ma.hw.ac.uk}

\maketitle
\thispagestyle{empty}

\abstracts{We consider Ising models defined on periodic approximants
of aperiodic graphs. The model contains only a single coupling
constant and no magnetic field, so the aperiodicity is entirely given
by the different local environments of neighbours in the aperiodic
graph. In this case, the partition function zeros in the temperature
variable, also known as the Fisher zeros, can be calculated by
diagonalisation of finite matrices.  We present the partition function
zero patterns for periodic approximants of the Penrose and the
Ammann-Beenker tiling, and derive precise estimates of the critical
temperatures.}

\section{Introduction}

The Ising model is the paradigm for a second-order phase transition in
two-dimensional models of statistical mechanics. Without an external
magnetic field, the partition function of the two-dimensional Ising
model can be calculated explicitly for regular lattices. However, if
the Ising model is defined on an aperiodic graph, for instance on the
Penrose tiling (PT), this is no longer the case. Therefore, the influence
of an aperiodic order in the underlying structure on the phase
transition of the Ising model has been investigated by various means,
including numerical simulations, series expansions, and zeros of the
partition function; an overview on the results and a comprehensive
lists of references on the subject can be found in recent review
articles.\cite{GB,G}

The most general prediction stems from heuristic scaling arguments,
adapted from a relevance criterion for disordered Ising
models.\cite{G} It yields an inequality involving a characteristic
exponent that describes the fluctuations in the disordered or
aperiodically ordered system, the correlation critical exponent $\nu$
of the pure system, and the space dimension of the fluctuation. Planar
quasiperiodic graphs obtained by cut-and-project methods\cite{GS} have
low fluctuations, because they are flat sections through
higher-dimensional periodic lattices. Therefore, the aperiodicity is
expected to be irrelevant in these cases, which conforms with the
results obtained for specific examples, notably the PT.

\begin{figure}[t]
\centerline{\epsfysize=5cm\epsfbox{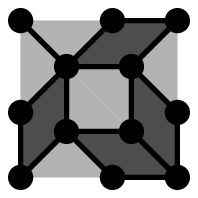}\hspace{1ex}
\epsfysize=5cm\epsfbox{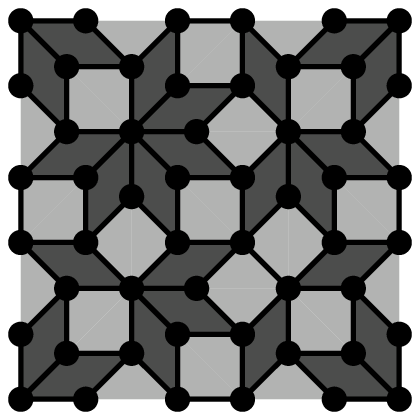}\hspace{1ex}
\epsfysize=5cm\epsfbox{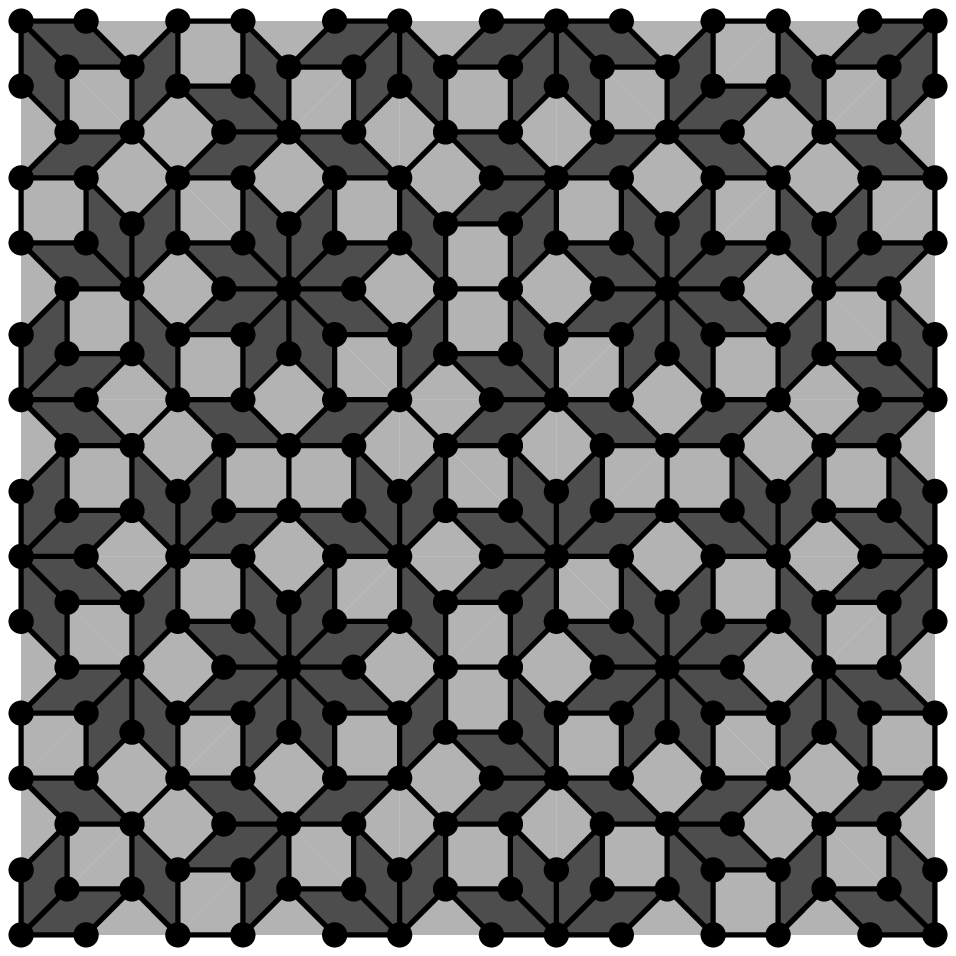}}
\caption{The first three periodic approximants 
($m=1,2,3$) of the Ammann-Beenker tiling.\label{fig:til}}
\end{figure}

Here, we consider partition function zeros obtained for periodic
approximants of the rhombic PT and the Ammann-Beenker tiling
(ABT). The unit cells of the first three approximants for the latter
are shown in Fig.~\ref{fig:til}.

\section{The Ising model}

We place Ising spins $\sigma_{j,k}^{}\in\{-1,1\}$ on the vertices of
the periodic approximants. Spins interact by a nearest-neighbour
interaction $J$. The energy of a configuration
$\sigma=\{\sigma_{j}^{}\mid 1\le j\le M\}\in\{-1,1\}^{M}$ on a finite
graph with $M$ vertices is
\begin{equation}
E(\sigma) = -\sum_{\langle j,k\rangle} J \sigma_{j}^{}\sigma_{k}^{},
\end{equation}
where the summation is performed over all pairs $\langle j,k\rangle$
of neighbouring spins, i.e., those located on vertices that are
connected by an edge. The corresponding partition function is the
sum on all configurations
\begin{equation}
Z(\beta) = \sum_{\sigma}\exp[-\beta E(\sigma)],
\end{equation}
where $\beta$ is the inverse temperature.  

\section{Partition function zeros}

We are interested in the pattern of zeros of this function in the
complex temperature plane.  In the thermodynamic limit
$M\rightarrow\infty$, the zeros accumulate, filling curves or areas in
the complex plane. These separate different regions of analyticity of
the free energy, which is essentially the logarithm of the partition
function, and hence correspond to phase transitions.

\begin{figure}[t]
\centerline{\epsfxsize=0.38\textwidth\epsfbox{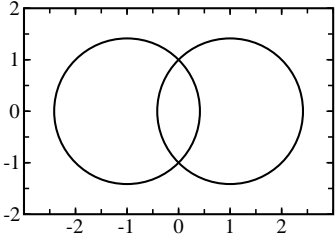}}
\caption{Partition function zeros of the square lattice Ising
model.\label{fig:sq}}
\end{figure}

The zero pattern for the square lattice case is shown in
Fig.~\ref{fig:sq}, which displays the zeros in the complex variable
$z=\exp(2\beta J)$. It consists of two circles. The two intersections
at $z=\sqrt{2}+1$ and $z=\sqrt{2}-1$ with the positive real axis
correspond to the ferromagnetic and antiferromagnetic critical points
$w_{\rm c}=\tanh(\beta_{\rm c} J)=\pm\sqrt{2}$, respectively.

\section{Results and conclusions}

The partition function zeros of the Ising model on any periodic planar
graph with $N$ spins in a unit cell can be calculated by diagonalising
a two-parameter family of $4N\times 4N$ matrices.\cite{RGS1,RGS2} The
powerful tool behind this method is the Kac-Ward determinant or
Pfaffian method for calculating the Ising model partition function on
general planar graphs.\cite{RGS1} Thus, this method allows to compute
partition function zeros for Ising models on infinite periodic graphs
with an arbitrary numerical precision.

\begin{figure}[t]
\centerline{\epsfbox{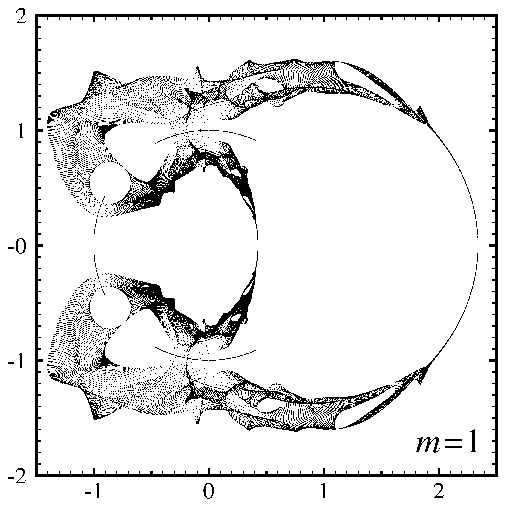}\hspace{1ex}\epsfbox{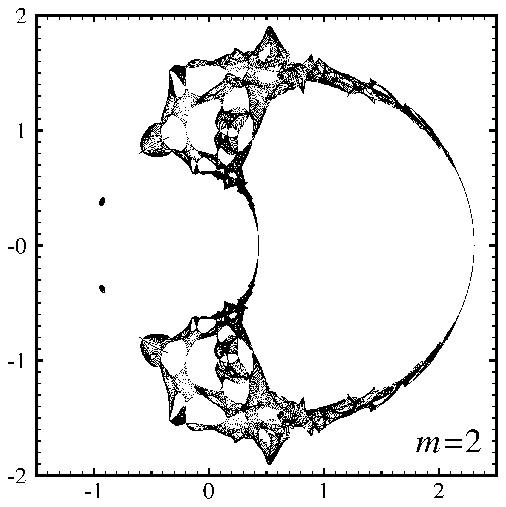}}
\centerline{\epsfbox{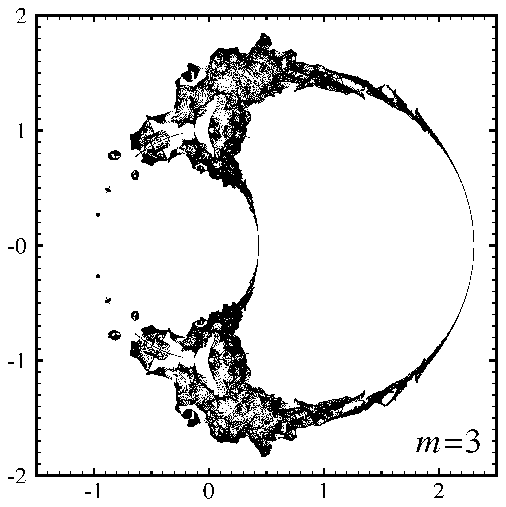}\hspace{1ex}\epsfbox{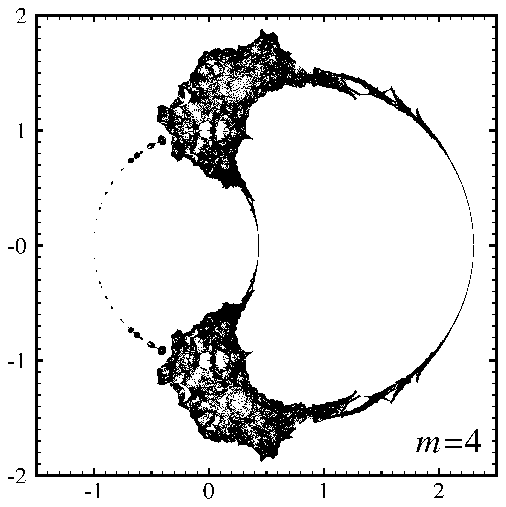}}
\centerline{\epsfbox{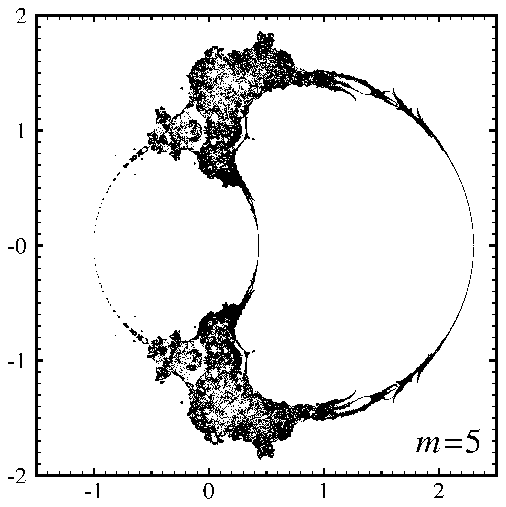}\hspace{1ex}\epsfbox{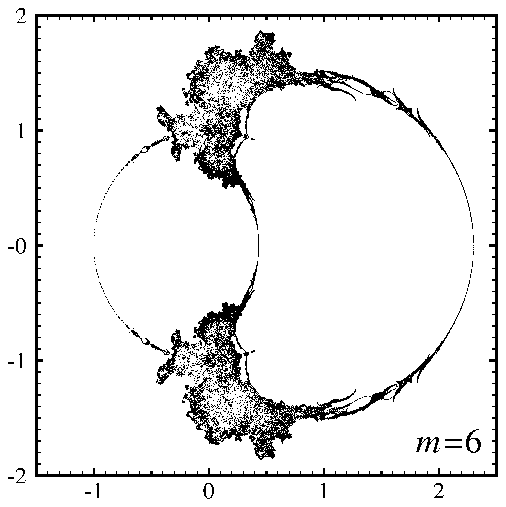}}
\caption{Part of partition function zeros for
periodic approximants of the Penrose tiling. \label{fig:pz}}
\end{figure}

\begin{figure}[t]
\centerline{\epsfbox{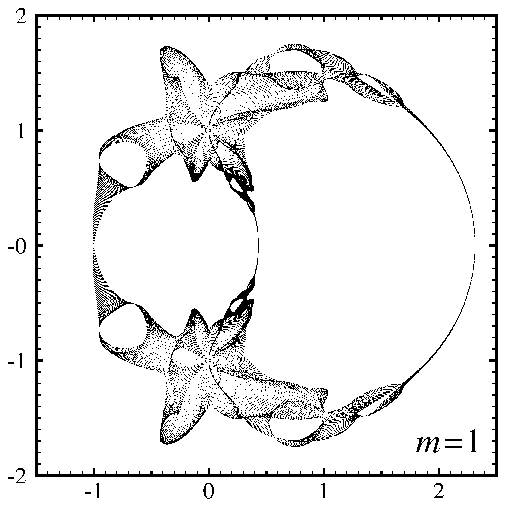}\hspace{1ex}\epsfbox{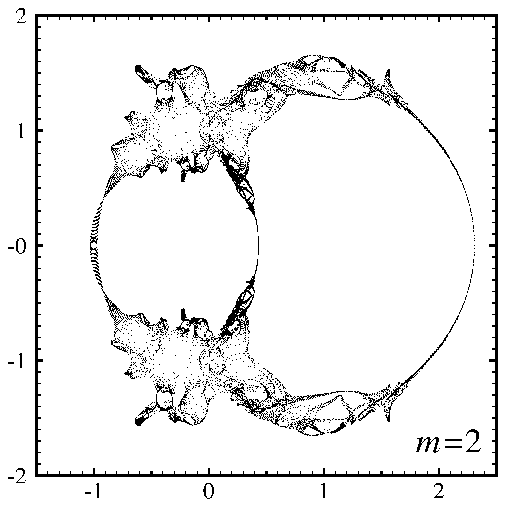}}
\centerline{\epsfbox{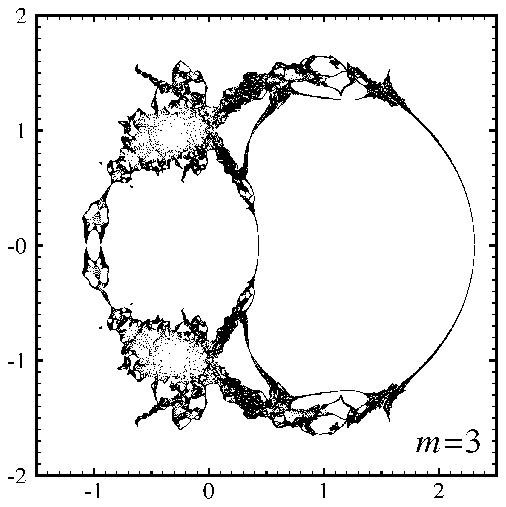}\hspace{1ex}\epsfbox{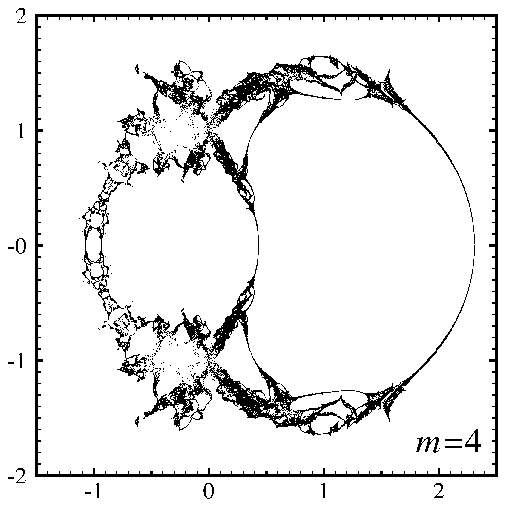}}
\caption{Partition function zeros for periodic approximants of
the Ammann-Beenker tiling. \label{fig:abz}}
\end{figure}

The result for the first approximants of the PT and the ABT are shown
in Figs.~\ref{fig:pz} and \ref{fig:abz}. These show accumulated zeros
obtained by diagonalising the Kac-Ward matrix for various values of
the two parameters, again in the complex variable $z=\exp(2\beta
J)$. The corresponding values of the critical temperature are listed
in Table~\ref{tab:ct}. These yield precise estimates for the critical
temperature for the aperiodic tilings, the errors were estimated on
the basis of the apparent convergence of the first few
terms.\clearpage

\clearpage 

The zero patterns, besides the trivial symmetry under complex
conjugation of $z$, still show the $z\leftrightarrow z^{-1}$ symmetry
that is a consequence of the bipartiteness of the tilings. This also
implies that the ferromagnetic and antiferromagnetic phase transitions
are of the same type. However, some of the symmetry of the square
lattice case, corresponding to the self-duality of the square-lattice
Ising model, is lost. In fact, the dual graphs of the aperiodic graphs
under consideration contain triangles, hence are not bipartite. The
antiferromagnetic model on these dual graphs will suffer frustration.
This may be responsible for the rather complicated, presumably fractal
or fractally bounded zero patterns, particularly further away from the
positive real axis. The zeros around the two phase transition points
still lie on a circle intersecting the real axis orthogonally,
consistent with the fact that the phase transition belongs to the same
universality class as on the square lattice.

\begin{table}[t]
\caption{Critical temperatures $w_{\rm c}=\tanh{(\beta_{\rm c}J)}$ for
periodic approximants of the PT and the ABT. Here, $m$ labels the
approximants with $N$ spins per unit cell.\label{tab:ct}}
\begin{center}
\addtolength{\tabcolsep}{1ex}
\begin{tabular}{|r|rl|rl|}
\hline
\rule[0ex]{0ex}{3ex}%
& \multicolumn{2}{c|}{{\bf \underline{Penrose tiling}}} &  
\multicolumn{2}{c|}{{\bf \underline{Ammann-Beenker tiling}}}\\
\rule[-1.5ex]{0ex}{1.5ex}%
\raisebox{2ex}{$m$} &
$N$ &
\multicolumn{1}{c|}{$w_{\rm c}$} &
$N$ &
\multicolumn{1}{c|}{$w_{\rm c}$} \\ \hline
\rule[0ex]{0ex}{3.25ex}%
$1$      &    $11$  & $0.401\, 440\, 380$ &
               $7$  & $0.396\, 850\, 570$ \\
$2$      &    $29$  & $0.395\, 411\, 099$ &
              $41$  & $0.396\, 003\, 524$ \\
$3$      &    $76$  & $0.395\, 082\, 894$ &
             $239$  & $0.395\, 985\, 346$ \\
$4$      &   $199$  & $0.394\, 554\, 945$ &
            $1393$  & $0.395\, 984\, 811$ \\
$5$      &   $521$  & $0.394\, 523\, 576$ &
            $8119$  & $0.395\, 984\, 795$ \\
$6$      &  $1364$  & $0.394\, 454\, 880$ &
           $47321$  & $0.395\, 984\, 795$ \\
$7$      &  $3571$  & $0.394\, 451\, 035$ & & \\
$8$      &  $9349$  & $0.394\, 441\, 450$ & & \\ 
$9$      & $24476$  & $0.394\, 439\, 826$ & & \\
\rule[-1.5ex]{0ex}{1.5ex}%
$10$     & $64079$  & $0.394\, 439\, 319$ & & \\ \hline   
\rule[-1.5ex]{0ex}{4.5ex}%
$\infty$ & $\infty$ & $0.394\, 439(1)$ & 
           $\infty$ & $0.395\, 984\, 79(1)$ \\ \hline
\end{tabular}
\end{center}
\end{table}

\end{document}